\documentclass[aps,prb,twocolumn,floatfix]{revtex4-1}
\usepackage{graphicx}
\usepackage{bm,amsmath,amssymb,mathrsfs,dcolumn}
\usepackage[usenames]{color}
\usepackage{subfigure}
\usepackage{ulem}

\begin{document}
%\title{\sout{Entanglement in quantum domain walls}\\
\title{Strong entanglement of spins inside a quantum domain wall}

\author{H. Y. Yuan}
\affiliation{Department of Physics, South University of Science and
Technology of China, Shenzhen 518055, Guangdong, China}

\author{Man-Hong Yung}
\email[Electronic address: ]{yung@sustc.edu.cn}
\affiliation{Institute for Quantum Science and Engineering and Department
of Physics, South University of Science and Technology of China, Shenzhen 518055,
Guangdong, China}

\author{X. R. Wang}
\email[Electronic address: ]{phxwan@ust.hk}
\affiliation{Department of Physics, The Hong Kong University of
Science and Technology, Clear Water Bay, Kowloon, Hong Kong}
\affiliation{HKUST Shenzhen Research Institute, Shenzhen 518057, China}
\date{\today}

\begin{abstract}
Magnetic domain walls (DWs) are widely regarded as classical objects
in physics community, even though the concepts of electron spins and
spin-spin exchange interaction are quantum mechanical in nature.
One intriguing question is whether DWs can survive at the quantum
level and acquire the quantum properties such as entanglement.
Here we show that spins within a DW are highly entangled in their
quantum description. The total magnetization of a magnetic DW is
nonzero, which is a manifestation of the global entanglement of the
collective spin state. These results significantly deepen our
understanding of magnetic DWs and enable the application of DWs in
quantum information science. The essential physics can be
generalized to skyrmions so that they can also play a role in
quantum information processing.
\end{abstract}

\maketitle
\textbf{Entanglement is a measure of quantum correlations between two or among
more than two quantum particles. It is a natural resource for quantum
computing and quantum information processing. Finding/generating,
extracting and utilizing this resourse is an important task in quantum
information science\cite{Nielsen}. Entangling tens particles have been realized
experimentally \cite{Neeley2010, DiCarlo2010,Barends2014,Bernien2017,
Zhang2017,Song2017}. Remarkably, in superconducting circuits and
trapped ions, it is now possible to control and to tune the coupling
strength between two spins from ferromagnetic to antiferromagnetic
interactions. These developments open intriguing possibilities for
studying the quantum properties of magnetic structures, such as DWs 
\cite{Hubert, Yuan2015,Yuan2016nonlocal}, vortices \cite{Shinjo2001,
Wachowiak2002,Choe2004,Yuan2015guide} and even skyrmions 
\cite{Bogdanov2001,Rosler2006,Muhlbauer2009, Yu2010, yuan2016}with a 
finite number of controllable spins.
To initiate this interdisciplinary field, it is essential to connect
magnetic structures with the measurable quantities of quantum information,
such as purity of spins and global entanglement of the system} \cite{yuan2017bec}.

Here we study a quantum magnetic wire with two magnetic domains and a DW
in between. We show that all spins in the DW are highly entangled.
The net magnetization of DWs does not depend on the nanowire length and
it can be well characterized by the global entanglement of the system.
In addition, we find that global entanglement of the system is a natural
indictor of the phase transition between quantum DWs and domains.

%\section{Model and Method}
%{\it\color{red}Model.---}
\vspace{6pt}

\noindent\textbf{Results}

\noindent\textbf{Model of quantum domain walls.} Let us consider the
transverse Ising model of $N$ spin-1/2 particles on a one dimensional
(1D) lattice, subject to boundary fields $h$, the Hamiltonian reads
\begin{equation}
\mathcal{H}=J\sum_{\langle ij \rangle} \sigma_i^z \cdot \sigma_j^z
- g\sum_{i=1}^N \sigma_{i}^x -h (\sigma_{1}^z + \sigma_{N}^z),
\label{ham}
\end{equation}
where $\sigma_i^x, \sigma_i^z$ are the Pauli matrices on the $i$-th
site, $J$ is the exchange coupling and $g$ is the anisotropy energy.
Here $\langle ij \rangle$ denotes nearest neighbor spins.
The ground-state energy $E_0$ and the corresponding ground state
$|0\rangle$ is calculated using the standard Lancoz algorithm
\cite{Lanczos1950,Dagotto1994} (See Methods for details), which is double checked by the exact
diagonization of the Hamiltonian $\mathcal{H}$ for $N \leq 8$.

In the following, we shall focus on the antiferromagnetic (AFM)
coupling  ($J>0$) for an illustration and discussion of our results.
The conclusions drawn below can be generalized to the ferromagnetic
coupling as well. In an AFM domain, the quantum version of the
magnetization order $m_i^z$ and N\'{e}el order $n_i^z$
\cite{Yuan2017direction} is
given by the following expectation values:
\begin{equation}
\begin{aligned}
m_i^z = \frac{1}{2} \langle 0| \sigma_i^z + \sigma_{i+1}^z |0\rangle,
n_i^z = \frac{1}{2} \langle 0| \sigma_i^z - \sigma_{i+1}^z |0\rangle, \\
\end{aligned}
\end{equation}
where $i=1,3,5,...N-1$. Without ambiguity, the number of spins is
assumed  even and $J=1.0$ if it is not stated otherwise.

%\section{Results and Discussions}

\begin{figure}
\centering
\includegraphics[width=0.5\textwidth]{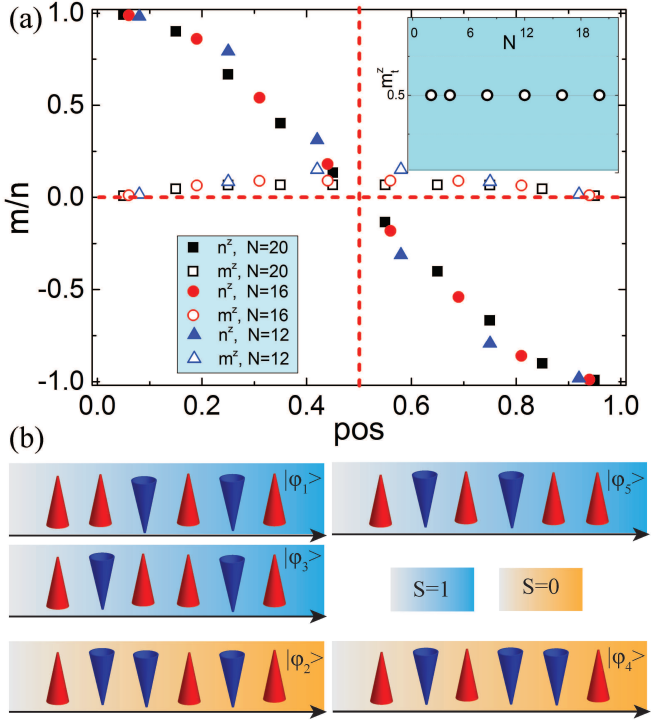}\\
\caption{\textbf{Scheme of quantum domain walls.} (a) Spatial variation
of $m^z$ (open symbols) and $n^z$ (filled symbols) for a spin chain
with $N=12$ (triangles), 16 (circles), 20 (squares), respectively.
The top-right inset shows the net magnetization as a function of $N$.
$\mathrm{pos} =i/N$ is the scaled position of the $i-$th site, $g=0.2.$
(b) Scheme of the five degenerated low energy excitation states for $N=6$.
The blue (orange) background represents the net magnetization of the
corresponding state to be $S=1$ ($S=0$).}
\label{fig1}
\end{figure}

%\subsection{Intrinsic magnetization}
%{\it\color{red}Intrinsic magnetization.---}
\vspace{6pt}
\noindent\textbf{Domain wall properties.} Let us first look at
the numerical results, in the regime with weak anisotropy $g \ll J$.
Figure \ref{fig1}a shows the spatial variation of $m_i^z$ and $n_i^z$
for $N=12$ (triangles), 16 (circles), 20 (squares), respectively.
These profiles are similar to those of classical AFM DWs. (1)
The N\'{e}el order $n_z$ has a typical classical DW profile whose
value varies from 1 on the left hand side (LHS) of the chain to $-1$
on the right hand side (RHS). (2) The magnetization order $m_z$ is
zero in the two domains and reaches its maximum at the DW center.
As the system becomes larger, the magnetic moments near the center
become smaller, but, by summing all the magnetic moments in the DW,
the net magnetization is $m_t^z = 1/2$, independent of the system
size ($N$) as shown in the inset of Fig. \ref{fig1}a.

To further prove the result of $m_t^z = 1/2$, we look at the structure
of a quantum DWs of $N$ spins. The boundary spins are aligned along
the $+z$ direction by the external fields. Given $g \ll J$, the AFM
exchange interaction dominates the Hamiltonian (\ref{ham}).
Thus, the lowest energy states (i.e. ground-state) are those states
with only one pair of neighboring spins left alone (sketched as a
pair of parallel spins in Fig. \ref{fig1}b) while rest nearest
neighboring spins are anti-parallel with each other forming a zero spin
singlet state. For $N=6$, there are five such configurations of same
energy as sketched in Fig. \ref{fig1}b. In general, there are $N-1$
such states of energy $-(N-3)J$. These states can be further classified
into $|\varphi^{S=1}_i \rangle$ for $S=1$ and $|\varphi^{S=0}_i \rangle$
for $S=0$, $N/2$ for $S=1$ and the remaining for $S=0$, where $S$ is the
net magnetization of the system.
The true ground state under the boundary conditions is the linear
combination of this states. The corresponding eigenstate of the
Hamiltonian (\ref{ham}) is
\begin{equation}
|0_{\mathrm{th}} \rangle = \sum_{i=1}^{N/2} a_i|\varphi^{S=1}_i \rangle
+\sum_{i=1}^{N/2-1} b_i|\varphi^{S=0}_i\rangle \
\end{equation}
where $a_i$,$b_i$ are the superposition coefficients.
By rearranging the basis states such that $S=1$ and $S=0$ states
are ordered alternatively, the Hamiltonian (\ref{ham}) can be recasted
as a tridiagonal-Toeplitz matrix, where the ground state energy is found
to be $E_0 = -(N-3)J - 2g \cos \left ( \pi / N\right )$ with the wave
function
\begin{equation}
|0_{\mathrm{th} \rangle} = \sqrt{\frac{2}{N}} \left ( \sin \frac{\pi}{N},
\sin \frac{2\pi}{N},..., \sin \frac{(N-1)\pi}{N} \right ),
\end{equation}
Note that the anisotropy term $g$ only gives a first-order correction
of the ground state energy. As a result, the magnetic moments distribution
is given by $m_i^z=\langle 0_{\mathrm{th}}| \sigma_i^z
+ \sigma_{i+1}^z |0_{\mathrm{th}}\rangle/2 = u_i$,
where $u_i = 2/N\sin^2 i\pi/N$. The net magnetization can be calculated as
\begin{equation}
m^z_t= \sum_{\mathrm{odd}~ i}^{N-1} u_i=\frac{1}{2},
\end{equation}
which does not depend on $N$, in agreement with numerical results.
The essential physics is that the ground state is a superposition of
total spin $S=1$ and $S=0$ configurations with equal contributions
($\sum u_{2i} = \sum u_{2i+1}=1/2$) so that the average spin number is always 1/2.

%\noindent\textbf{Chiral domain walls.}
In fact, the above result is also true when the Dzyaloshinsiki-Moriya interaction
(DMI) \cite{Dz1957, Moriya1960} $H_{\mathrm{DM}}=-\mathbf{D} \cdot
\sum_{\langle ij\rangle} \sigma_i \times \sigma_j$ is added to Hamiltonian
(\ref{ham}), where the ground state of the revised Hamiltonian is still
solvable analytically. Specifically, for $\mathbf{D}=D\hat{y}\ (D>0)$,
the ground state is given by,
\begin{equation}
\begin{aligned}
|0_{\mathrm{th}}\rangle = \sqrt{\frac{2}{N}} \left ( \tau_1 \sin \frac{\pi}{N},
\tau_2 \sin \frac{2\pi}{N},...,
\tau_{N-1}\sin \frac{(N-1)\pi}{N} \right ),
\end{aligned}
\end{equation}
where the prefactor $\tau_i =1-2\ \mathrm{mod} ( \lceil i/2 \rceil,2)$,
$\lceil x \rceil$ is the ceiling function of $x$.
Note that $\tau_i^2=1$, the magnetization distribution
$m_i^z = 2/N \sin^2 (i\pi/N)$, is not changed by the DMI.
However, the chirality of DW does depend on the sign of $D$.
For $D>0$, the system always prefers a clockwise N\'{e}el wall.
However, if $D$ reverses the sign, the ground state becomes
counter-clockwise. This is consistent with classical case \cite{yuan2017}.

%\subsection{Purity and global entanglement}
%{\it\color{red}Entanglement.---}
\vspace{6pt}
\noindent\textbf{Entanglement.} In quantum information science, the
purity of the $i$-th spin is defined as $P_i =\mathrm{tr}(\rho_i^2)$
where $\rho_i$ is the density matrix of the $i$-th spin obtained by
tracing all the other spins in the full density matrix $|0\rangle\langle0|$.
It quantifies the distance of a state relative to a pure state; for example,
it takes the value 1 for a separable pure state, but 1/2 for a maximally
entangled states such as Bell states and Greenberg-Horne-Zelinger state
(GHZ) states \cite{GHZ2007}.

\begin{figure}
\centering
  % Requires \usepackage{graphicx}
\includegraphics[width=0.5\textwidth]{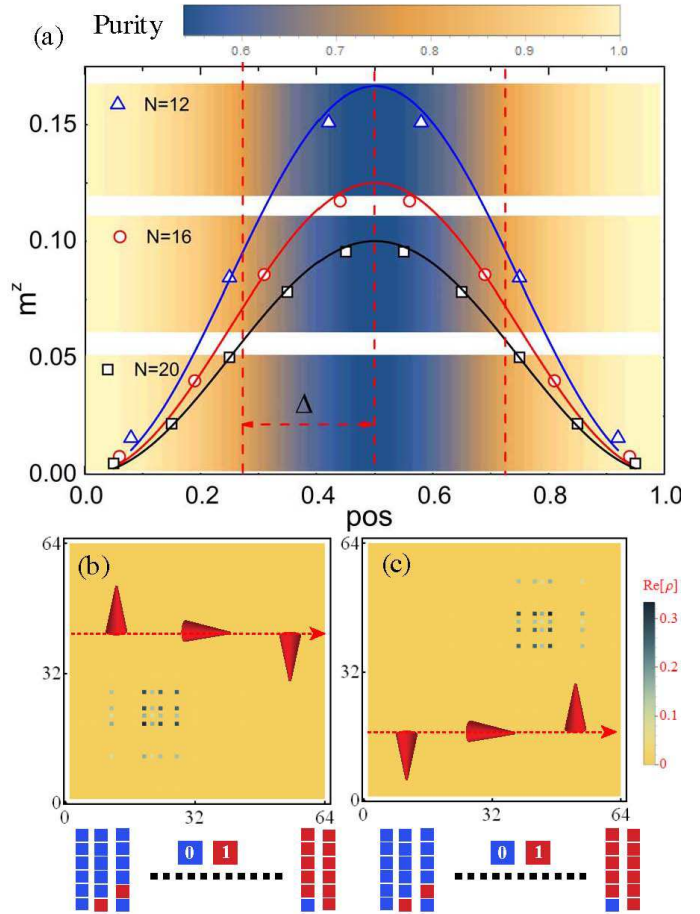}\\
\caption{\textbf{Tomography of quantum domain walls.}
(a) Distribution of the magnetic moments of DWs for $N=12$ (triangles),
$16$ (circles) and $20$ (squares), respectively. The solid lines are
the theoretical results given by $m_i^z =u_i$. The colored strips
represent the distribution of purity at the corresponding $N$.
(b) (c) Density of states tomography for a clockwise/counter-clockwise
DW state in a spin chain with $N=6$. The values of vertical/horizontal
axis represents the value of the basis vector by treating up spin as
bit 0 and down spin as bit 1. The inset sketches the DW profile.}
\label{fig2}
\end{figure}

Figure \ref{fig2}a shows the space distribution of purity as colored
rectangles for $N=12$ (top inset), $16$ (middle inset), and $20$ (bottom inset),
respectively. Regardless of the system size, the purity takes 1 near
the boundary of spin chain and then decreases to form a symmetric dip
around the chain center at $i/N=1/2$. The magnetization distribution
of DWs takes on a peak centered at the dip position and its space
dependence is strongly correlated with the purity. Physically, this
correlation originates from the antiferromagnetic nature of nearest
neighbour interaction. For the spins near the boundary, their directions
are strongly bounded by the fixed orientations of the boundary spins
and have a larger probability to be in N\'{e}el-state-like configurations.
Consequently, the magnetization is close to zero and these spins are not
entangled with the others (i.e., close to a pure state).
The spins near the center are influenced lightly by the boundary spins
and their directions become much more uncertain, hence the purity of
these spins become smaller. The purity of spins close 1/2 at the wire center
 suggests that the spins inside the domain walls are highly entangled.

Theoretically, the purity of the $i-$th spin is defined by the
reduced density operator of the  $i-$th spin obtained by tracing
all the spins except the $i-$th spin in density operator
$|0_{\mathrm{th}} \rangle \langle 0_{\mathrm{th}} |$ as
\begin{equation}
P_i^{\mathrm{th}} =\left (\sum_{k=1}^{i-1} u_k \right )^2
+ \left(\sum_{k=i}^{N-1} u_k \right)^2 + 2u_i u_{i-1},
\label{purity}
\end{equation}
where $i=1,2,...,N$. The third term $u_i u_{i-1}$ is of the order
$O(1/N^2)$, which is much smaller than the first two terms.
Considering $\sum u_i=1$, $P_i$ reaches its minimum value
(1/2) at $i = N/2$, i.e. the chain center, where
$m_i^z = u_i \propto \sin^2 (i\pi/N)$ reaches its maximum value.

The correlations of the magnetization distribution and purity distribution
can be extended to two dimensions (2D) where rich magnetic structures
such as vortices, skyrmions can exist since the physics remain the same.
Nevertheless, it is a great challenge to find the ground states in 2D
both analytically and numerically. %As a preliminary extension, we
%arrange the 1D spin chain with DMI in an axle-like way (See the
%Supplemental Materials for details) and plot the magnetization
%distribution and purity distribution in Supplement. Both distributions
%takes on a clear skyrmion profile.
Practically, the correlations also allow us to measure a quantum DW, vortices
and even skyrmions by measuring the purity of each spin in a spin chain. 
The later can be realized by performing state tomography on each system 
qubit or by a more efficient technique that uses bitwise interactions 
between the system and identically prepared registers \cite{Brennen2003}.
A typical tomography of the density of states for a clockwise/counter-clockwise
DW is shown in Fig. \ref{fig2}b and \ref{fig2}c. The distinguishable
distributions of the density matrix allows us to classify clockwise and
counter-clockwise DWs.

Furthermore, the direct sum of magnetization recovers the net magnetization
of the spin chain while the proper average of the local purity can
give the global entanglement ($E_g$) of the spins. The original definition
of $E_g$ is given by Meyer and Wallach \cite{Meyer2002} and then
reformulated in terms of the sum of local purity \cite{Brennen2003}, i.e.

\begin{equation}
E_g \equiv \frac{2}{N} \sum_{i=1}^N (1-P_i).
\end{equation}
Substituting Eq. (\ref{purity}) into this definition, the leading order
of global entanglement is reduced to
\begin{equation}
E_g = 2 - \frac{2}{N} \sum_{k=1}^N \left [ \left(\sum_{i=1}^{k-1} u_i \right)^2
+ \left(\sum_{i=k}^{N-1} u_i \right )^2 \right ].
\end{equation}
The sum increases as $N$ increases and saturates for $N>20$. The limiting
value is $E_g = 2/3-5/2\pi^2 \approx 0.41$. This indicates that the
spins in a DW is still entangled in the macroscopic scale
($N\rightarrow \infty $). This is quite different from a classic DW
that has zero entanglement since all its composite spins have definite
orientations.

As a comparison, for a quantum domain state without DWs such as the
superposition of two degenerated N\'{e}el states,
\begin{equation}
|0_{\mathrm{th}} \rangle = \frac{1}{\sqrt{2}}\left (|\uparrow \downarrow \uparrow 
\downarrow...\uparrow \downarrow \rangle
+ \right |\downarrow \uparrow \downarrow \uparrow ...\downarrow \uparrow \rangle)
\label{neel}
\end{equation}
The global entanglement is $E_g = 1$ while its net magnetization is zero.
Based on this comparison, it becomes possible to distinguish domain and
DW states and the corresponding net magnetization by measuring the
global entanglement of the spin system.

\begin{figure}
\centering
  % Requires \usepackage{graphicx}
\includegraphics[width=0.5\textwidth]{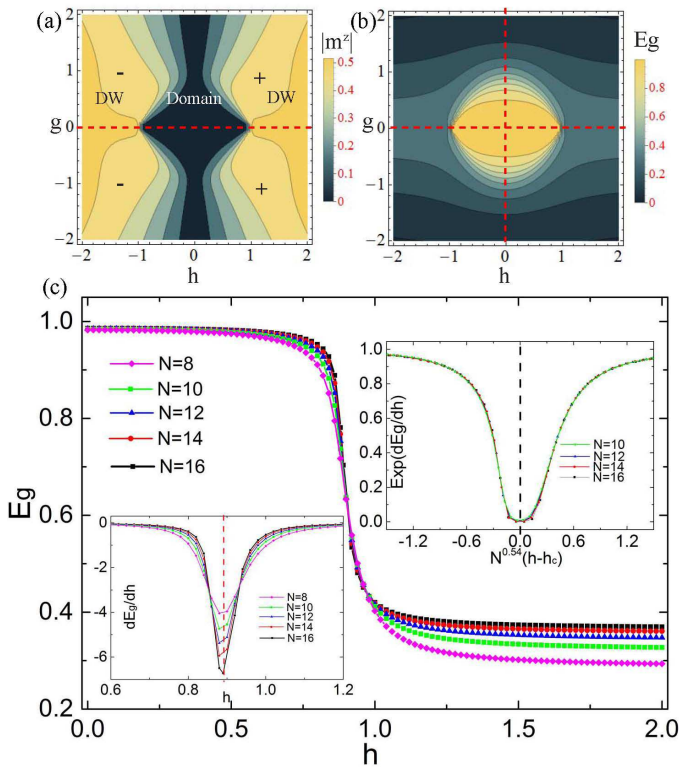}\\
\caption{\textbf{Phase diagram and the characterization of the phase
transition using entanglement.} (a) The distribution of net magnetization
of the spin chain ($|\mathrm{m}^z|$) in the $h - g$ phase space.
The $\pm$ signs represent clockwise and counter-clockwise DW respectively.
$N=12$.
(b) Simulated global entanglement ($E_g$) distribution in the $h - g$ phase space.
(c) Global entanglement ($E_g$) as a function of boundary field ($h$). $g = 0.2$.
Left inset: $dE_g/dh$ as a function of external field for
$N=8,10,12,14,16$, respectively. Right inset: scaling behavior of the
global entanglement around the critical point $h_c$. }
\label{fig3}
\end{figure}
%\subsection{Phase transition}
%{\it\color{red}Phase transition.---}
%\vspace{6pt}
%\noindent\textbf{Temperature dependence of global entanglement.}
%To consider the influence
%of temperature on the global entanglement of the system, both the ground state and excited
%states have to be considered.
%
%Figure \ref{fig3} shows the global entanglement distribution in the $k_BT/J \sim N$ plane.
%Given a particular $N$, the global entanglement first increases with temperature and then
%saturates around 2/3. For a typical magnetic material, $k_B/J = 600$ K, then the figure
%clearly indicates the existence of entanglement at the room temperature ($T=300$ K).
%This may be due to the topological stability of the quantum DWs against the temperature.

\vspace{6pt}
\noindent\textbf{Phase transition.} To know the size of the phase space
where the quantum DWs are stable, it is interesting to plot the phase
diagram of the spin model. In our numerical simulation, we fix the
exchange coupling and adjust the anisotropy $g$ and boundary fields to
obtain various ground states in $h-g$ plane as shown in Fig. \ref{fig3}.

Guided by the red line in Fig. \ref{fig3}a, the ground state changes
from a single domain state to a quantum DW at the critical field of
$h_c = J$ for $g=0$. The critical field has a non-trivial
dependence on $g$ and $h$. For simplicity, we focus on the regime
$g \ll 1$, where the ground state is mainly determined by the
competition of exchange interaction and Zeeman interaction.
In this regime, the global entanglement ($E_g$) shows an abrupt
jump from 1.0 to a smaller value as shown in \ref{fig3}b and
Fig. \ref{fig3}c with $g=0.2$. This abrupt change is due to the
distinguishable quantum properties of the domains and DWs.
When $h <h_c$, the exchange interaction dominates the Hamiltonian
and the ground state is a superposition of two degenerated N\'{e}el
states Eq. (\ref{neel}) with $E_g=1$. The numerical value of $E_g$
is a little smaller than the theoretical value 1, as shown in
Fig. \ref{fig3}c, due to the influence of the small anisotropy $g$
term, which tendency is to align all the spins to $x$ direction and
reduce the entanglement. As the field increases above $h_c$,
the ground state becomes a DW state with finite entanglement 0.41
as discussed previously.

The global entanglement and its first derivative is continuous with $h$
near the critical field while its second order derivative is discontinuous,
as shown in the left inset of Fig. \ref{fig3}c. To verify whether this
discontinuity can be a good indictor of the phase transition, we first
plot the first derivative of $E_g$ versus $h$ in the left inset of Fig.
\ref{fig3}c for system size ranging from $N=8, 10, 12, 14,16$.
As $N$ increases, $dE_g/dh$ shows a clear divergence tendency at the critical field.
To eliminate the finite size effect, we do scaling analysis for finite systems
using the scaling ansatz \cite{Osterloh2002},
\begin{equation}
\frac{dE_g}{dh}=\ln N^\nu |h-h_c|,
\end{equation}
where $\nu$ is the critical exponent. As shown in the right inset of
Fig. \ref{fig3}c, $\nu =0.54$ gives perfect scaling results for the
finite systems. This consistency shows that global entanglement is a
good measure of quantum DW/domain phase transition in this system.

%\section{Conclusions}
%{\it\color{red}Conclusions.---}
\vspace{6pt}
\noindent\textbf{Discussion}

\noindent We have shown that an AFM DW in a 1D lattice has an intrinsic
magnetization of 1/2 independent of lattice length. The reason is that
the ground state is a superposition of $S=1$ states and $S=0$ states
of equal amplitude. The global entanglement of such a DW is non-zero
and even exists at a macroscopic scale. Moreover, the magnetization
profile is closely related to the local purity of spins such that the
DW width can be extracted from the purity profile that can be measured
through state tomography in quantum information science.
The typical energy gap of the ground state and the first excited state
in our model is $43 ~\mu\mathrm{eV}$ ($\sim 500$ mK) for $N=20$.
Then the mixture of excited states and ground state can be neglected
at a temperature below 500 mK, which leaves a sufficient room to
experimentally verify our theoretical prediction.

Since our results are applicable to chiral DWs due to DMI, the main
conclusions can be generalized to two dimensional case for magnetic
skyrmions. This type of skyrmions has non-zero entanglement that is
very different from the classical skyrmions discussed in literature.
Further study of the quantum skyrmions may lead to
quantum skyrmion spintronics.
%\section{Acknowledgement}

\vspace{6pt}
\noindent\textbf{Methods}

\noindent\textbf{Lanczos algorithm.}
Our numerical calculation of the ground states of the spin model are
based on the Lanczos method. First, a random initial state
$| \psi_0 \rangle$ is chosen in the $2^N$ dimensional Hilbert space.
Then we define a new state $| \psi_1 \rangle$ that is orthogonal to
$| \psi_0 \rangle$, realized by subtracting
$\mathcal{H} | \psi_0 \rangle$ over $| \psi_0 \rangle$ i.e.

\begin{equation}
| \psi_1 \rangle = \mathcal{H} | \psi_0 \rangle -  a_0| \psi_0 \rangle
\end{equation}
where $a_0 = \langle \psi_0 | \mathcal{H} | \psi_0 \rangle / \langle \psi_0|\psi_0 \rangle$.
Next, we define the state $|\psi_2\rangle $ that is orthogonal to both $|\psi_0 \rangle $ and
$|\psi_1 \rangle $,
\begin{equation}
| \psi_2 \rangle = \mathcal{H} | \psi_1 \rangle - a_1 | \psi_1 \rangle - b_0 | \psi_0 \rangle
\end{equation}
where $a_1 = \langle \psi_1 | \mathcal{H} | \psi_1 \rangle / \langle \psi_1|\psi_1 \rangle$,
$b_0 = \langle \psi_1|\psi_1 \rangle /\langle \psi_0|\psi_0 \rangle$.
Through iterations, the state $|\psi_{n+1} \rangle$ is derived as
\begin{equation}
| \psi_{n+1} \rangle = \mathcal{H} | \psi_n \rangle - a_n | \psi_n \rangle - b_{n-1} | \psi_{n-1} \rangle,
\end{equation}
where the coefficients $a_n = \langle \psi_n | \mathcal{H} | \psi_n \rangle / \langle \psi_n|\psi_n \rangle$,
$b_{n-1} = \langle \psi_n|\psi_n \rangle /\langle \psi_{n-1}|\psi_{n-1} \rangle$.

In terms of the mutual orthogonal basis spanned by $\{|\psi_0\rangle, |\psi_1\rangle,...|\psi_n\rangle \}$,
the Hamiltonian can be written in the form of a tridiagonal matrix
and then diagonalized through the standard subroutine.
The ground state energy $E_n$ is obtained as the smallest eigenvalues
of the Hamiltonian. As $n$ increases, the $E_n$ will converge to the
real ground state energy of the system while the corresponding eigen-vector
is the wave function of ground states. The convergence criteria used in the
simulations is $|E_n - E_{n-1}| \leq 10^{-14}$.

\vspace{6pt}
%\noindent \textbf{References}

\vspace{6pt}
\noindent\textbf{Acknowledgements}

\noindent
H.Y.Y. acknowledges the financial support from National Natural Science
Foundation of China (NSFC) Grant (No. 61704071).
MHY acknowledges support by NSFC Grant (No. 11405093), Guangdong
Innovative and Entrepreneurial Research Team Program (2016ZT06D348),
and Science, Technology and Innovation Commission of Shenzhen
Municipality (ZDSYS20170303165926217 and JCYJ20170412152620376).
XRW was supported by the NSFC Grant (No. 11774296) as well
as Hong Kong RGC Grants (Nos. 16300117 and 16301816).

\vspace{6pt}
\noindent \textbf{Author contributions}

\noindent H.Y.Y., M.H.Y. and X.R.W conceived the project. H.Y.Y. performed
the numerical and theoretical calculations.  All authors
analyzed and discussed the results and contributed to the writing of the
manuscript.

\vspace{6pt}
\noindent \textbf{Additional information}

\noindent The authors declare no competing financial interests.

%\newpage
%\noindent \textbf{Supplementary Material}

\begin{thebibliography}{}

%\bibitem{MacDonald2008} P. M. Haney
%and A. H. MacDonald, Phys. Rev. Lett. {\bf 100}, 196801 (2008).
%
%\bibitem{Ke2008} Y. Xu, S. Wang, and K. Xia, Phys. Rev. Lett.
%{\bf 100}, 226602 (2008).
%
%\bibitem{Jungwirth2011} T. Jungwirth, V. Nov\'{a}k, X. Marti, M. Cukr,
%F. M\'{a}ca, A. B. Shick, J. Ma\u{s}ek, P. Horodysk\'{a}, P. N\u{e}mec,
%V. Hol\'{y}, J. Zemek, P. Ku\u{z}el, I. N\u{e}mec, B. L. Gallagher,
%R. P. Campion, C. T. Foxon, and J. Wunderlich, Phys. Rev. B
%\textbf{83}, 035321 (2011) and references therein.
%
%\bibitem{Hals2011} K. M. D. Hals, Y. Tserkovnyak, and A. Brataas,
%Phys. Rev. Lett. \textbf{106}, 107206 (2011).
%
%\bibitem{Cheng2014} R. Cheng, J. Xiao, Q. Niu, and A. Brataas,
%Phys. Rev. Lett. \textbf{113}, 057601 (2014).
%
%\bibitem{Barker2016} J. Barker and O. A. Tretiakov, Phys. Rev. Lett.
%\textbf{116}, 147203 (2016).
%
%\bibitem{Gomonay2016} O. Gomonay, T. Jungwirth, and J. Sinova,
%Phys. Rev. Lett. \textbf{117}, 017202 (2016).
%
%\bibitem{Wadley2016} P. Wadley et al., Science \textbf{351}, 587 (2016).
%
%\bibitem{Yuan2017} H. Y. Yuan and X. R. Wang, Appl. Phys. Lett.
%\textbf{110}, 082403 (2017).
%
%\bibitem{Shiino2016} T. Shiino, S. Oh, P. M. Haney, S. -W. Lee, G. Go,
%B. -G. Park, and K. -J. Lee, Phys. Rev. Lett. \textbf{117}, 087203 (2016).
%
%\bibitem{Tveten2016} E. G. Tveten, T. Muller, J. Linder, and A. Brataas,
%Phys. Rev. B \textbf{93}, 104408 (2016).
%
%\bibitem{Horodecki2009} R. Horodecki, P. Horodecki, M. Horodecki,
%and K. Horodecki, Rev. Mod. Phys. \textbf{81}, 865 (2009).
%
\bibitem{Nielsen} M. A. Nielsen and I. L. Chuang, \textit{Quantum Computation
and Quantum Information}, 10th Anniversary Ed. (Cambridge Unviersity Press, 2000).

\bibitem{Neeley2010} Neeley, M. et al. Generation of three-qubit entangled states
using superconducting phase qubits. \textit{Nature} \textbf{467}, 570-573 (2010).

\bibitem{DiCarlo2010} Dicarlo, L. et al. Preaparation and measurement of
three-qubit entanglement in a superconducting circuit. \textit{Nature} \textbf{467}, 574-578 (2010).

\bibitem{Barends2014} Barends, B. et al. Superconducting quantum circuits
at the surface code threshold for fault tolerance. \textit{Nature} \textbf{508}, 500-503 (2014).

\bibitem{Bernien2017} Bernien, H. et al. Probing many-body dynamics on a 51-atom quantum simulator.
\textit{Nature} \textbf{551}, 579-584 (2017).

\bibitem{Zhang2017} Zhang, J. et al. Observation of a many-body dynamical phase transition
with a 53-qubit quantum simulator. \textit{Nature} \textbf{551}, 601-604 (2017).

\bibitem{Song2017} Song, C. et al. 10-qubit entanglement and parallel logic operations with
a superconducting circuit. \textit{Phys. Rev. Lett.} \textbf{119}, 180511 (2017).

%\bibitem{Ladd2010} T. D. Ladd, F. Jelezko, R. Laflamme, Y. Nakamura, C. Monroe,
%and J. L. O'Brien, Nature \textbf{464}, 45 (2010).
%
%%\bibitem{Simon1994} D.R.Simon,
%
%\bibitem{Clarke2008} J. Clarke and F. K. Wilhelm, Nature \textbf{453}, 1031 (2008).
%
%\bibitem{Imamog1999} A. Imamog-lu, D. D. Awschalom, G. Burkard, D. P. DiVincenzo, D. Loss,
%M. Sherwin, and A. Smail, Phys. Rev. Lett. \textbf{83}, 4204 (1999).
%
%\bibitem{Weber2010} J. R. Weber, W. F. Koehl, J. B. Varley, A. Janotti,
%B. B. Brckley, C. G. Van de Walle, and D. D. Awschalom,
%PNAS \textbf{107}, 8513 (2010).
%
%\bibitem{Dagotto1994} E. Dagotto, Rev. Mod. Phys. \textbf{66}, 763 (1994).
%
%\bibitem{Brennen2003} G. K. Brennen, Quant. Inf. and Comput. \textbf{3}, 616 (2003).
%
%\bibitem{Meyer2002} D. A. Meyer and N. R. Wallach, J. Math. Phys., \textbf{43}, 4273 (2002).
%
%\bibitem{Osterloh2002} A. Osterloh, L. Amico, G. Falci, and R. Fazio,
%Nature \textbf{416}, 608 (2002).
%
%\bibitem{Cooper1956} Cooper, L. N. Bound electron pairs in a degenerate Fermi gas.
%\textit{Phys. Rev.} \textbf{104}, 1189-1190 (1956).

%\bibitem{MacDonald2008} Haney, P. M.
%\& MacDonald, A. H. Current-induced torques due to compensated antiferromagnets.
%\textit{Phys. Rev. Lett.} {\bf 100}, 196801 (2008).
%
%\bibitem{Ke2008} Xu, Y., Wang S. \& Xia, K. Spin-transfer torques in antiferromagetic
%metals from first principles. \textit{Phys. Rev. Lett.}
%{\bf 100}, 226602 (2008).
%
%\bibitem{Jungwirth2011} Jungwirth, T. et al., \textit{Phys. Rev. B}
%\textbf{83}, 035321 (2011) and references therein.
%
%\bibitem{Hals2011} Hals, K. M. D., Tserkovnyak, Y. \& Brataas, A.,
%Phenomenology of current-induced dynamics in antiferromagnets.
%\textit{Phys. Rev. Lett.} \textbf{106}, 107206 (2011).
%
%\bibitem{Cheng2014} Cheng, R., Xiao, J., Niu, Q. \& Brataas, A.
%Spin pumping and spin-transfer torques in antiferromagnets.
%\textit{Phys. Rev. Lett.} \textbf{113}, 057601 (2014).
%
%\bibitem{Barker2016} Barker J. \& Tretiakov, O. A. Static and dynamic
%properties of antiferroamgnetic skyrmions in the presence of
%applied current and temperature. \textit{Phys. Rev. Lett.}
%\textbf{116}, 147203 (2016).
%
%\bibitem{Gomonay2016} Gomonay, O., Jungwirth, T. \& Sinova, J.
%High antiferromagnetic domain wall velocity induced by N\'{ee}l
%spin-orbit torques.
%\textit{Phys. Rev. Lett.} \textbf{117}, 017202 (2016).
%
%\bibitem{Wadley2016} Wadley P. et al., Electrical switching of
%an antiferromagnet. \textit{Science} \textbf{351}, 587-590 (2016).
%
%\bibitem{Yuan2017} Yuan H. Y. \& Wang, X. R. Magnon-photon coupling
%in antiferromagnets. \textit{Appl. Phys. Lett.}
%\textbf{110}, 082403 (2017).
%
%\bibitem{Shiino2016} Shiino, T. et al. Antiferromagetic domain wall
% motion driven by spin-orbit torques. \textit{Phys. Rev. Lett.} \textbf{117}, 087203 (2016).
%
%\bibitem{Tveten2016} Tveten, E. G., Muller, T., Linder, J., \& Brataas, A.
%Intrinsic magnetization of antiferromagnetic textures.
%\textit{Phys. Rev. B} \textbf{93}, 104408 (2016).
\bibitem{Hubert} Hubert, A. \& Sch\'{a}fer, R. \textit{Magnetic domains: the analysis of magnetic
microstructures.} (Springer Berlin Heidelberg 1998).

\bibitem{Yuan2015} Yuan, H. Y. \& Wang, X. R. Boosing domain wall propagation
by notches. \textit{Phys. Rev. B} \textbf{92}, 054419 (2015).

\bibitem{Yuan2016nonlocal} Yuan, H. Y., Yuan, Z., Xia, K., \& Wang, X. R. Influence of nonlocal damping
on the field-driven domain wall motion. \textit{Phys. Rev. B}  \textbf{94}, 064415 (2016).

\bibitem{Shinjo2001} Shinjo, T. et al. Magnetic vortex core observation
in circular dots of permalloy.
\textit{Science} \textbf{289}, 930-932 (2001).

\bibitem{Wachowiak2002} Wachowiak, A. et al. Direct observation of internal spin structure of
magnetic vortex cores. \textit{Science} \textbf{298}, 577-580 (2002).

\bibitem{Choe2004} Choe, S.-B. et al. Vortex core-driven magnetization dynamics.
\textit{Science} \textbf{304}, 422 (2004).

\bibitem{Yuan2015guide} Yuan, H. Y. \& Wang, X. R. Nano magnetic vortex wall guide. \textit{AIP Advances},
\textbf{5}, 117104 (2015).\\

\bibitem{Bogdanov2001} Bogdanov, A. N. \& R\"{o}{\ss}ler, U. K.
Chiral symmetry breaking in magnetic thin films and multilayers.
\textit{Phys. Rev. Lett.} \textbf{87}, 037203 (2001).

\bibitem{Rosler2006} R\"{o}{\ss}ler, U. K. \& Bogdanov, A. N. \& Pfleiderer, C.
Spontaneous skyrmion ground states in magnetic metals.
\textit{Nature(London)} \textbf{442}, 797-801 (2006).

\bibitem{Muhlbauer2009} M\"{u}hlbauer, S. et al. Skyrmion lattice in a chiral magnet.
\textit{Science} \textbf{323}, 915-919 (2009).

\bibitem{Yu2010} Yu, X. Z. et al. Real-space observation of a two-dimensional skyrmion
crystal. \textit{Nature(London)} \textbf{465}, 901-904 (2010).

\bibitem{yuan2016} Yuan, H. Y. \& Wang, X. R. Skyrmion creation and manipulation by
nano-second current pulses.
\textit{Sci. Rep.} \textbf{6}, 22638 (2016).

\bibitem{yuan2017bec} Yuan, H. Y. \& Yung, M. -H. Thermal entanglement of magnonic condensates.
Preprint at arXiv: 1711.04394.

%\bibitem{Horodecki2009} Horodecki, R., Horodecki, P., Horodecki, M.
%and Horodecki, K. Quantum entanglement. \textit{Rev. Mod. Phys.} \textbf{81}, 865-942 (2009).
%
%\bibitem{Nielsen} Nielsen M. A. and Chuang, I. L. \textit{Quantum Computation
%and Quantum Information}, 10th Anniversary Ed. (Cambridge Unviersity Press, 2000).
%
%\bibitem{Ladd2010} Ladd, T. D. et al. Quantum computers. \textit{Nature} \textbf{464}, 45-53 (2010).
%
%%\bibitem{Simon1994} D.R.Simon,
%
%\bibitem{Clarke2008} Clarke J. \& Wilhelm, F. K. Superconducting quatnum bits.
%\textit{Nature} \textbf{453}, 1031-1042 (2008).
%
%\bibitem{Imamog1999} Imamog-lu, A. et al. Quantum information processing using quatnum
%dot spins and cavity QED. \textit{Phys. Rev. Lett.} \textbf{83}, 4204 (1999).
%
%\bibitem{Weber2010} Weber, J. R. et al. Quantum computing with defects.
%PNAS \textbf{107}, 8513-8518 (2010).

\bibitem{Lanczos1950} Lanczos, C. An iteration method for the solution of the eigenvalue
problems of linear differential and integral operators. \textit{J. Res. Nat. Bur. Stand.}
\textbf{45}, 255-282 (1950).

\bibitem{Dagotto1994} Dagotto, E. Correlated electrons in high-temperature supreconductors.
\textit{Rev. Mod. Phys.} \textbf{66}, 763-840 (1994).

\bibitem{Yuan2017direction} Yuan, H. Y., Wang, W., Yung, M. -H. \& Wang, X. R.
Classification of amgnetic forces on antiferromagnetic domain wall. Preprint at
arXiv:1712.03055.

\bibitem{Dz1957} Dzyaloshinskii, I. E. Thermodynamic theory of weak ferromagnetization
in antiferromagnetic substances.
\textit{Sov. Phys. JETP} \textbf{5}, 1259 (1957).

\bibitem{Moriya1960} Moriya, T. Anisotropic superexchange interaction and weak
ferromagnetism. \textit{Phys. Rev.} \textbf{120}, 91-98 (1960).

\bibitem{yuan2017} Yuan, H. Y., Gomonay, O. \& Kl\"{a}ui, M. Skyrmions and multi-sublattice
helical states in a frustrated chiral magnet. \textit{Phys. Rev. B} \textbf{96}, 134415 (2017).

\bibitem{GHZ2007} Greenberger, D. M., Horne, M. A., \& Zeilinger, A. Going beyond Bell's
theorem, Preprint at arXiv:0712.0921 (2007).

\bibitem{Brennen2003} Brennen, G. K. An observable measure of entanglement for pure states
of multi-qubit systems. \textit{Quant. Inf. and Comput.} \textbf{3}, 616 (2003).

\bibitem{Meyer2002} Meyer D. A. \& Wallach, N. R. Global entanglement in multiparticle
systems. \textit{J. Math. Phys.} \textbf{43}, 4273-4278 (2002).

\bibitem{Osterloh2002} Osterloh, A., Amico, L., Falci, G., \& Fazio, R.,
Scaling of entanglement close to a quantum phase transition. \textit{Nature} \textbf{416}, 608-610 (2002).
\end{thebibliography}
\end{document}